\documentclass[%
 reprint,
superscriptaddress,
 amsmath,amssymb,
 aps,
pra
]{revtex4-2}

\usepackage{xcolor}
\usepackage{graphicx}
\usepackage{dcolumn}
\usepackage{bm}
\usepackage{siunitx}
\begin{document}
\title{Interplay of plasmonics and strain for Hexagonal Boron Nitride emission engineering}
\author{Anuj Kumar Singh}
\thanks{These two authors contributed equally}
\author{Utkarsh}
\thanks{These two authors contributed equally}
\affiliation{%
Laboratory of Optics of Quantum Materials, Department of Physics, Indian Institute of Technology Bombay, Mumbai- 400076, India 
}%
\author{
Pablo Tieben}
\affiliation{Leibniz University Hannover, Germany}
\author{Kishor Kumar Mandal}
\author{Brijesh Kumar}
\affiliation{%
Laboratory of Optics of Quantum Materials, Department of Physics, Indian Institute of Technology Bombay, Mumbai- 400076, India 
}%
\author{Rishabh Vij}
\affiliation{Department of Condensed Matter Physics and Material Science, Tata Institute of Fundamental Research, Homi Bhabha Road, Mumbai, 400005 India}
\author{Amrita Majumder}
\author{Ikshvaku Shyam}
\author{Shagun Kumar}
\affiliation{%
Laboratory of Optics of Quantum Materials, Department of Physics, Indian Institute of Technology Bombay, Mumbai- 400076, India 
}%
\author{Kenji Watanabe}
\affiliation{Research Center for Electronic and Optical Materials, National Institute for Materials Science, 1-1 Namiki, Tsukuba 305-0044, Japan}
\author{Takashi Taniguchi}
\affiliation{Research Center for Materials Nanoarchitectonics, National Institute for Materials Science,  1-1 Namiki, Tsukuba 305-0044, Japan}
\author{Venu Gopal Achanta}
\affiliation{Department of Condensed Matter Physics and Material Science, Tata Institute of Fundamental Research, Homi Bhabha Road, Mumbai, 400005 India}
\author{Andreas Schell}
\affiliation{Leibniz University Hannover, Germany}
\author{Anshuman Kumar}%
 \email{anshuman.kumar@iitb.ac.in}
\affiliation{%
Laboratory of Optics of Quantum Materials, Department of Physics, Indian Institute of Technology Bombay, Mumbai- 400076, India 
}%
\affiliation{Centre of Excellence in Quantum Information, Computation, Science and Technology, Indian Institute of Technology Bombay, Powai, Mumbai- 400076, India}
\date{\today}
\begin{abstract}
In the realm of quantum information and sensing, there has been substantial interest in the single-photon emission associated with defects in hexagonal boron nitride (hBN). With the goal of producing deterministic emission centers, in this work, we present a platform for engineering emission in hBN integrated with gold truncated nanocone structures. Our findings highlights that, the activation of emission is due to the truncated gold nanocones. Furthermore, we measure the quantum characteristics of this emission and find that while our system demonstrates support for single-photon emission, the origin of this emission remains ambiguous. Specifically, it is unclear whether the emission arises from defects generated by the induced strain or from alternative defect mechanisms. This uncertainty stems from the fluorescence properties inherent to gold, complicating our definitive attribution of the quantum emission source. To provide a rigorous theoretical foundation, we elucidate the effects of strain via the Kirchhoff-Love theory. Additionally, the enhancements observed due to plasmonic effects are comprehensively explained through the resolution of Maxwell's equations. This study will be useful for the development of deterministic and tunable single photonic sources in two dimensional materials and their integration with plasmonic platforms. 
\end{abstract}
\maketitle

\section{Introduction}
The deterministic generation of stable, controllable, and high-purity, bright quantum emitters is essential for the development of quantum photonic technology\cite{Moody2022,Gao2023, Gottscholl2021,Aichele2004}. So far, the generation of solid-state quantum emitters and engineering their emission have been investigated in several optical materials such as silicon carbide (SiC) \cite{lienhard2016bright, wang2018bright}, silicon nitride (SiN) \cite{senichev2021room, senichev2022silicon, smith2020single, mandal2024emission}, 2D transition metal dichalcogenides\cite {koperski2015single, parto2022cavity, azzam2021prospects}, hexagonal boron nitride (hBN) \cite{elshaari2021deterministic, froch2020coupling}, NV centers in diamond \cite{hausmann2012integrated, aharonovich2011diamond, mouradian2015scalable}, layered gallium selenide (GaSe) \cite{luo2023deterministic, tonndorf2017single}, and moire-trapped excitons \cite{baek2020highly}, and quantum dots of III-V semiconductors \cite{katsumi2019quantum, zadeh2016deterministic, davanco2017heterogeneous}. Epitaxially grown quantum dots, ultra-cold atoms and optically active defects in various materials provide a wide range of possibilities for Single Photon Emission (SPE) applications \cite{Arakawa2020, Bao2012, deOliveira2014}.
Defect states in van der Waals materials, in particular, have generated excitement due to their ease of integrability with existing on-chip photonic platforms\cite{MichaelisdeVasconcellos2022, Li2021, kumar2023photonic}.

\begin{figure*}
    \includegraphics[width=1.03\textwidth]{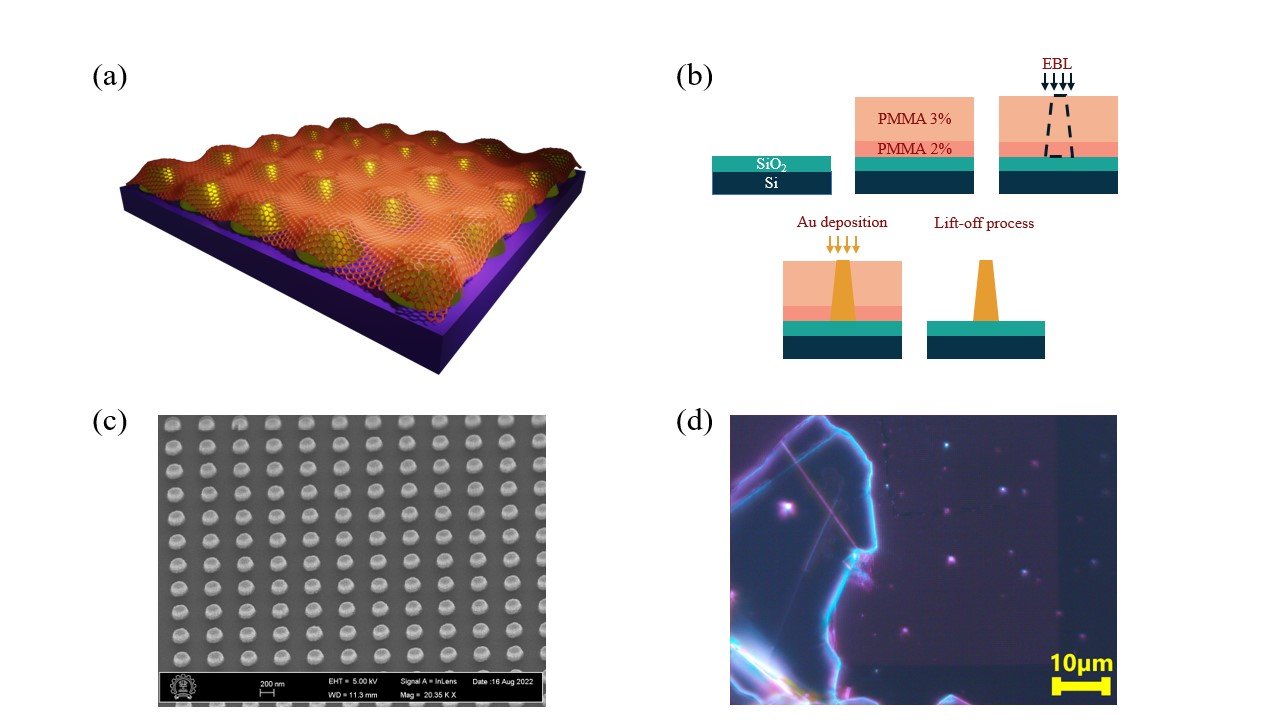}
    \caption{\textbf{Fabrication of hBN integrated plasmonic truncated nanocone structures.} (a) Schematic of hBN integrated  truncated nanocone structures (b) Schematic showing the fabrication steps of gold cone array by using a bi-layer resist. (c) SEM image of the truncated nanocone array with inset presenting the dimensions of the fabricated truncated nanocones. (d)  A false-colored dark-field image of transferred hBN flake on a gold truncated nanocone array clearly distinguishing the truncated nanocone region. A PDMS assisted dry transfer technique was employed here.}
    \label{fab}
\end{figure*}
\begin{figure*}
    \includegraphics[width=1.03\textwidth]{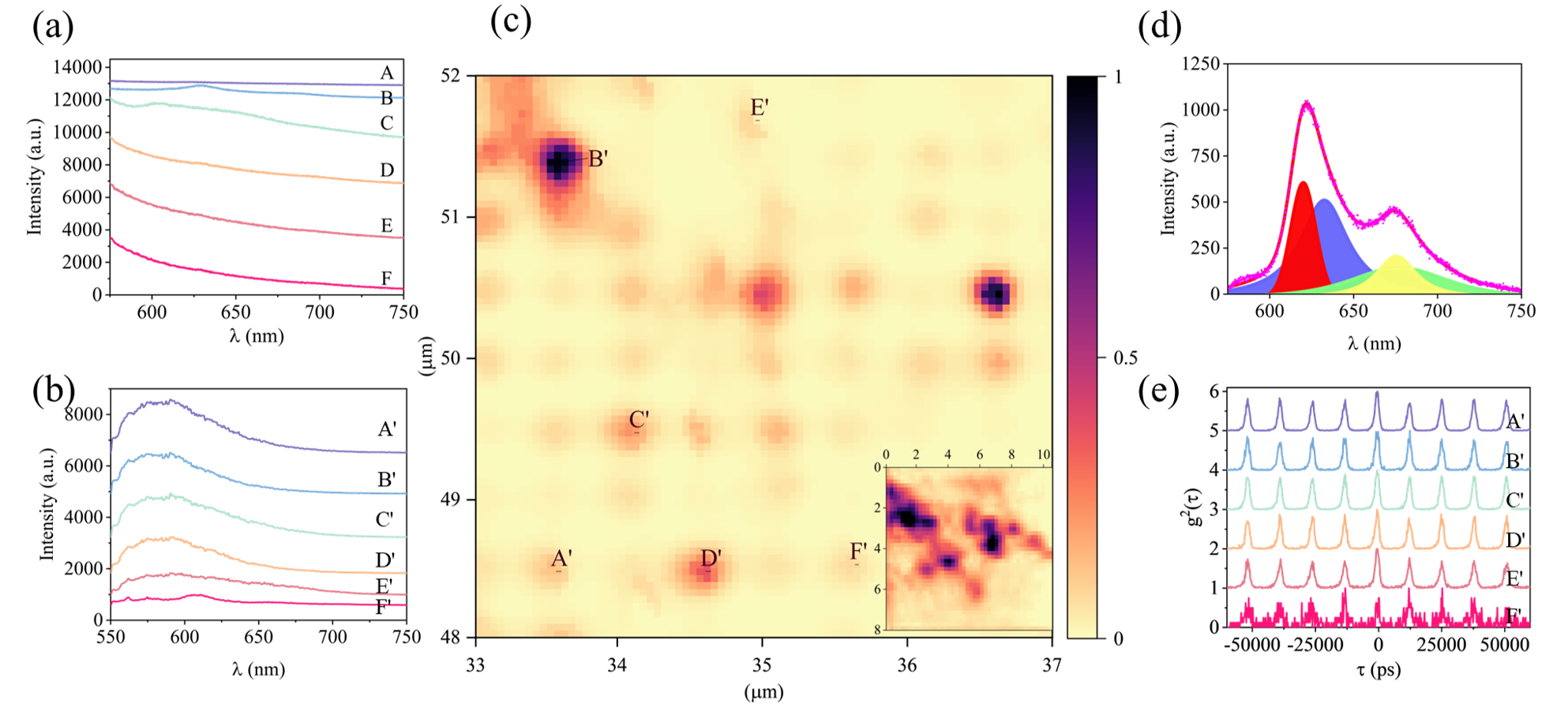}
    \caption{\textbf{hBN photoluminescence comparison on bare Si and gold truncated nanocones.} (a) Representative PL spectra from few layer hBN on Si/SiO$_2$. (b) Representative PL spectra from few layer hBN on truncated nanocones (c) Confocal PL intensity map at excitation power of $\SI{250}{\micro W}$ of few layer hBN ( for further confirmation that emission is coming from the hBN a coarse scan is taken at the 1.1 mW laser excitation power shown in  inset) (d) Representative PL spectra from hBN on gold truncated nanocones fitted for double ZPL ($\sim$620.07 nm and $\sim$632.58 nm) and PSB peaks also observed in previous studies \cite{Bommer2019}. (e) Measured $g^{(2)}(\tau)$ for the points shown in (c).}
    \label{pl-color}
\end{figure*}
\begin{figure*}
    \includegraphics[width=1.03\textwidth]{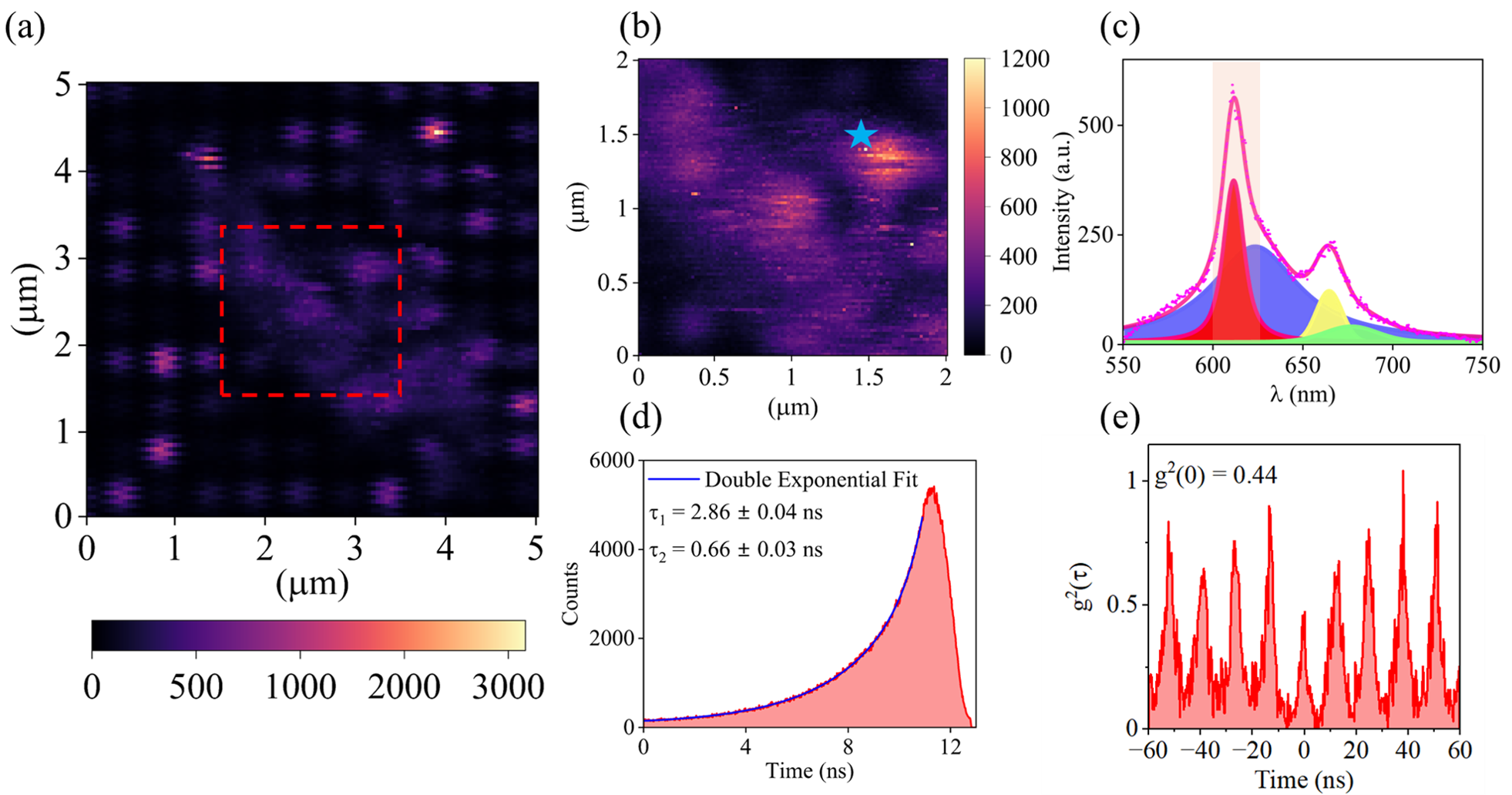}
    \caption{\textbf{Single photon emitter characterisation.} (a) Confocal intensity map of the few layer hBN over truncated nanocones. Gold truncated nanocone emission is clearly visible along with localised emission from hBN on top. (b) Finer confocal intensity map of hBN over gold truncated nanocones. (c) PL spectrum of an observed SPE clearly showing ZPL and PSB contribution. Highlighted region specifies the spectrum used for auto-correlation measurements. (d) Lifetime measurement at the specified point with a double exponential fit matching observed lifetimes for hBN SPEs \cite{Patel2022}, \cite{Bommer2019} (e) Second-order auto-correlation measurement with an observed $g^{(2)}(0)=0.44$.}
    \label{g2}
\end{figure*}
Hexagonal Boron Nitride (hBN), a van der Waals material, has emerged as a valuable photonic platform hosting room temperature-stable quantum emitters\cite{Grosso2017}. Typically, these defect states remain dormant due to the absence of charge carriers at the defect energies. Activation methods include the application of strain, coupling with plasmonics, e-beam irradiation, and polar liquids \cite{Chen2023,Xu2022,Singh2023,Kumar2023, Ronceray2023}. This enables the creation of defects that emit at desired wavelengths, highly tunable with the application of strain and proximity effects, owing to the van der Waals nature of the crystal.
Strain tuning of the emission wavelength has been explored via various routes such as hBN monolayers transferred onto a  hole array, arrays of dielectric and plasmonic nano-cylinders  and wrinkling of hBN\cite{Guo2023, Yim2020, Proscia2018, Proscia2019}. These SPEs are often accompanied by phonon side bands (PSB), representing the photoluminescence of phonon-assisted radiative decay.\\
In this work, we utilized a gold (Au) truncated plasmonic nanocones array fabricated via e-beam lithography to activate defects within hBN. Our findings reveal that hBN does not exhibit emission in the absence of integration with truncated Au cone structures. Furthermore, we delve into the quantum behavior exhibited by these emitters. However, it remains inconclusive whether the single-photon emitters are activated by the strain induced by the truncated nanocone structures or by alternative mechanisms. This report underscores the dual role of gold -- while it promotes emission and it also inhibits the quantum emission capabilities inherent to hBN.

\section{Methods}
\subsection{Fabrication: Plasmonic Au nanocone array}

A large-scale ($\SI{100}{\micro\metre}\times \SI{100}{\micro\metre}$) and relatively high aspect ratio Au-plasmonic truncated nanocone array was fabricated using a high-resolution electron beam lithography (EBL, RAITH150 Two system) and electron beam evaporator tools. A 300 nm thin silicon dioxide (SiO$_{2}$) bottom cladding layer was thermally grown (in an oxidation furnace at $1050 ~^{\circ} \mathrm{C}$ temperature) on RCA cleaned $\SI{275}{\micro\metre}$ thick commercially available silicon wafer (Si-$\langle 110\rangle$ substrate). The patterning process began with the cleaning of SiO$_{2}$/Si sample with Acetone/IPA/DI water followed by drying it with N$_{2}$ blow gun, UV Ozone cleaner (Ossila) treatment and heating on a hot plate. A stack of high-resolution and process-stable PMMA (Poly methyl methacrylate) e-beam resist of different sensitivity (950K-A2 and 950K-A3) was implemented to obtain resist thickness relatively higher than deposited Au. Firstly, the sample was coated with PMMA-950K-A2 resist at a moderate spin rate of  3000 revolutions per minute (RPM) followed by soft baking at $180 ~^{\circ} \mathrm{C}$ on a hot plate for 2 min. After that, three consecutive layers of PMMA-950K-A4 were spin-coated at the same conditions.

Subsequently, the e-beam exposure parameters to write the desired mask (through conventional area mode of writing) on the $\SI{100}{\micro\metre}$  write field area were set as 8~mm working distance, 30kV accelerating voltage, $\SI{10}{\micro\metre}$ aperture size, 0.044 nA irradiated focused e-beam current, 500 C/cm$^{2}$ area dose at 1000x magnification in $4.2\times10^{-6}$ mbar chamber vacuum. The exposed resist was developed on MIBK: IPA (mixture ratio 1:3) developer solution for 75 sec and rinsed in IPA at room temperature followed by drying with N$_{2}$ blow. Au metal of 400 nm thickness (with 10~nm Ti as an adhesion layer) was deposited on the EBL-processed sample using an e-beam evaporator at a constant slow deposition rate in a chamber base vacuum $\sim$10$^{-8}$ mbar. The Au-deposited sample was kept on PG remover at $80 ~^{\circ} \mathrm{C}$ on a hot plate for 12 hr for the lift-off process to remove the unpatterned resist area. Lastly, the Au-truncated nanocone plasmonic array structure is left on the SiO$_{2}$/Si substrate after successfully removing the resist. At the end, the sample is exposed to oxygen plasma to remove residual resin on the Au-truncated nanocone for vertical sidewall smoothening, as shown in Fig.~\ref{fab}(b).
\subsection{Integration: hBN on truncated Au cone array}
The hBN flakes were first exfoliated using Nitto blue tape on homemade polydimethylsiloxane (PDMS) film by mixing the PDMS solution and curing agent in a 10:1 ratio and then kept overnight at room temperature. Finally, these flakes were integrated atop of truncated nanocone structure using the deterministic home-built dry transfer setup, optical image as illustrated in Fig.~\ref{fab}(d). 
\subsection{Optical characterization setup}
We performed a comparative photoluminescence (PL) study at a 532 nm excitation wavelength for the hBN integrated with the truncated nanocone. All the PL studies were carried out using the Invia Reflex micro-Raman spectrometer with the help of a 100x microscope objective with NA $\sim$0.90. Raman spectrometer was initially calibrated with standard Raman peak of crystalline silicon at 520.7 cm$^{-1}$. To further study the quantum characteristics of the emission, second-order correlation function ($g^{(2)}(\tau)$) measurements deploying a free space Hanbury Brown Twiss (HBT) setup were performed. 

\section{Results and Discussion}
The fabrication process flow of Au truncated nanocones nanophotonic structure and the integration of hBN is illustrated in Fig.~\ref{fab}, as discussed in the last section. Additionally, a tilted scanning electron microscopy (SEM) image is presented in Fig.~\ref{fab}(c). Finally, the optical microscopic image of the integrated hBN of thickness of approximately tens of nanometer with truncated nanocone structures is shown in Fig.~\ref{fab}(d).
\begin{figure*}
    \includegraphics[width=0.8\textwidth]{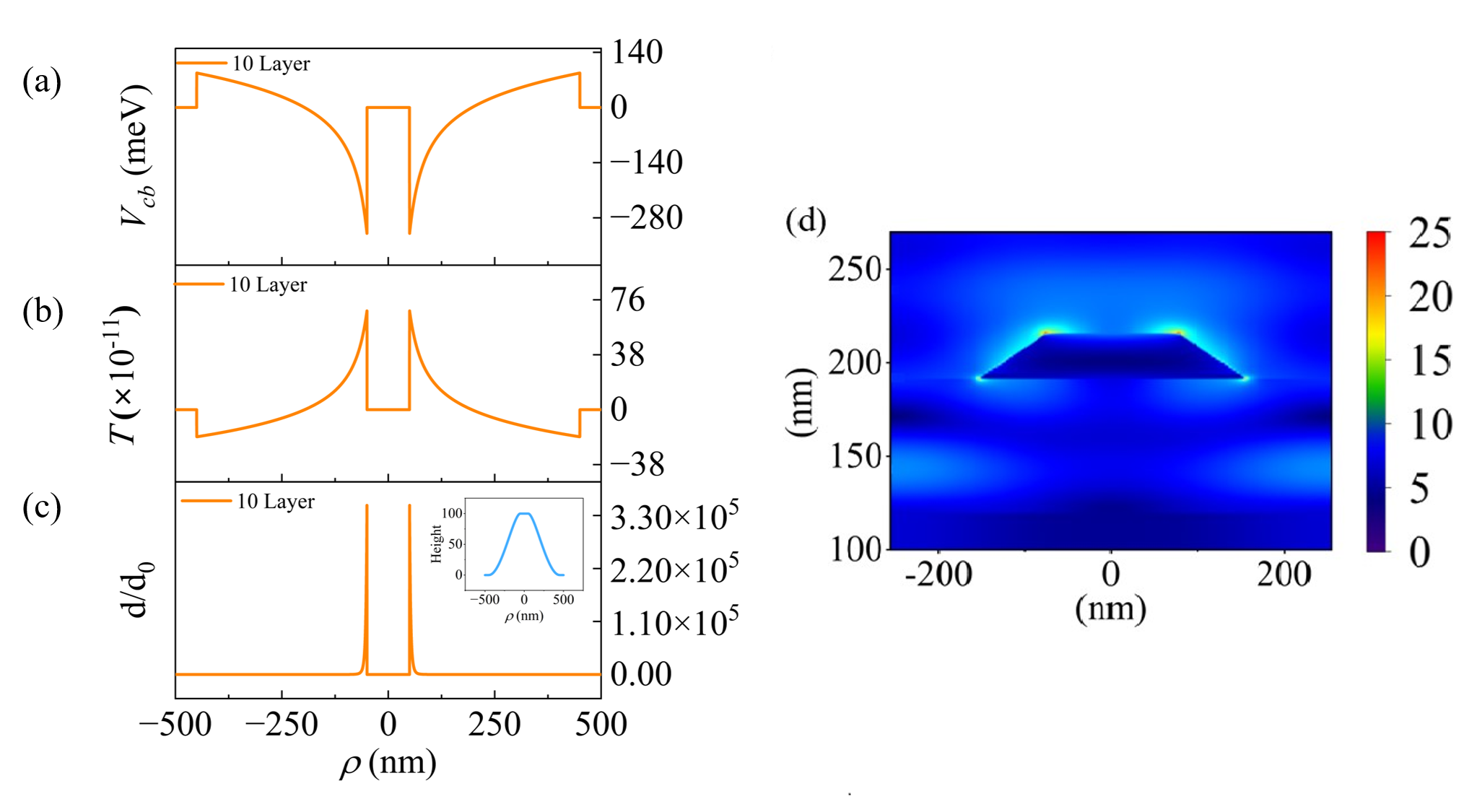}
    \caption{\textbf{Strain tuning and plasmonic enhancement.} (a) Deformation potential for the conduction band (b), (c) Corresponding strain and electronic charge density profile with inset for the height profile of hBN on a truncated nanocone. (d) FDTD simulation of electric field profile cross-section about the gold truncated nanocone showing plasmonic enhancement around the corners of the  truncated nanocone structure.}
    \label{strain}
\end{figure*}

The fabricated hBN-truncated nanocone hybrid system underwent photoluminescence (PL) spectroscopy with a 532 nm excitation wavelength. Fig.~\ref{pl-color}(a) displays the PL spectra of hBN on a bare Si/SiO$_2$ substrate, showing no significant emission peaks. This observation indicates the high purity of almost defect-free hBN. Subsequently, PL spectroscopy was conducted on hBN integrated with truncated nanocone samples, as depicted in Fig.~\ref{pl-color}(b) and extracted from the positions shown in the confocal PL map in Fig.~\ref{pl-color}(c) taken at an excitation power of $\SI{250}{\micro W}$. In the inset of Fig.~\ref{pl-color}(c), the coarse map of PL intensity at high laser power (1.1 mW) is plotted. The erratic counts in the PL map suggest potential fluorescence originating from the Au truncated nanocones\cite{Huang2019}, \cite{Pniakowska2022}. We show a representative PL spectrum from this map (excitation power of 1.1 mW) in Fig.~\ref{pl-color}(d). This spectra was well fitted using four voigt peaks, in Fig.~\ref{pl-color}(d), revealing two zero-phonon lines (ZPLs) at 620.07 nm and 632.58 nm and two phonon sidebands (PSBs) at 675.58 nm and 677.17 nm. Furthermore, in Fig.~\ref{pl-color}(e), we plot the measured $g^{(2)}$($\tau$) for various representative points from the map Fig.~\ref{pl-color}(c). These second-order correlation measurements were performed at an average excitation power of $\SI{250}{\micro W}$. From these results, it is confirmed that most of the emission is not quantum, which can be attributed to the fluorescence of the truncated Au nanocones. This observation makes it evident that the presence of gold fluorescence may have a detrimental impact on the observation of single photon emission from this platform.

To further investigate the quantum emission characteristics in our system, we conducted confocal photoluminescence (PL) scans. It is noteworthy that emission from the gold truncated nanocones can also be observed in the confocal scan shown in Fig.~\ref{g2}(a). The presence of gold fluorescence can interfere with the measurement of quantum behaviour originating from the hBN, as demonstrated in Fig.~\ref{pl-color}. A more detailed scan of a smaller $\SI{2}{\micro\metre}\times \SI{2}{\micro\metre}$ area (dashed red lines in Fig.~\ref{g2}(a)) was performed, as shown in Fig.~\ref{g2}(b). A specific emitter located at the position $(\SI{1.5}{\micro\metre},\SI{1.5}{\micro\metre})$ (marked by a star) was identified within this area, exhibiting quantum behavior at room temperature. The photoluminescence spectrum of this emitter, presented in Fig.~\ref{g2}(c), and fitted by four voigt peaks, revealed two zero-phonon line (ZPL) at 611.65 nm, 623.56 nm and two phonon side-band (PSBb) at 664.78 nm and 677.09 nm similar to the higher excitation power spectra presented in Fig.~\ref{pl-color}(d). The shift in the peak positions might be due to the varrying strains in different positions provided by the nanocone \cite{Mendelson2020}.

The time-resolved response of the emitter, as shown in Fig.~\ref{g2}(d), exhibited an exponential fluorescence decay with durations of approximately 2.86$\pm$0.04 \cite{Patel2022}ns and 0.66$\pm$0.03 ns \cite{Bommer2019} . In Fig.~\ref{g2}(e), the second-order correlation function $g^{(2)}$($\tau$) for the shaded pink region of Fig.~\ref{g2}(c) was collected using a spectral filter. The observed dip in $g^{(2)}$(0) of 0.44 confirms the quantum behaviour of the emitter. The quantum emission may arise from either a strained defect within the hBN or an edge defect at the hBN boundary. It is essential to note that the data presented here has not been corrected for background effects. Since we did not use a spacer layer between the metal and hBN, there is a possibility of non-radiative decay of emission states which corresponds to the observed $g^{(2)}(0)>$0.5 for most of the hBN emission. However, the presence of a single-photon emitter with gold fluorescence suggests that there might be a number of SPEs in the system which remain undetected due to such non-radiative pathways.

We modeled the height profile (Fig.~\ref{strain} (c) inset) of a thin hBN crystal on top of a 100 nm high, inner radius of 100 nm and outer radius of  150 nm applying Kirchoff-Love thin plate theory \cite{Sortino2020,Brooks2018}. Following the case of uniaxial strain , we obtained the corresponding strain profile as the trace of the strain tensor as shown in Fig.~\ref{strain}(b):
\begin{equation}
T=Tr[\epsilon_{ij}]=\frac{(2\sigma-1)h}{1-\sigma}\Delta\zeta   \end{equation}
 where $\sigma$ is the Poisson’s ratio, h is the thickness of the hBN, $\zeta$ and is the height profile function of the hBN layer. The strain has extremes at the edge of the truncated nanocones and the interaction point of hBN on SiO$_2$. Following a tight-binding approach \cite{Sortino2020,Brooks2018}, we calculate the deformation potential at the $K(K')$ point of the Brillouin zone. It is proportional to the strain profile as presented in Fig.~\ref{strain}(a).
\begin{equation}
V=\begin{pmatrix}
V_{vb}&0\\
0&V_{cb}
\end{pmatrix}=\begin{pmatrix}
\delta_v\mathcal{T}&0\\
0&\delta_c\mathcal{T}
\end{pmatrix}
\end{equation}
Here, V$_{vb}$ and V$_{cb}$ are the valence and conduction band potential, $\delta_v$ and $\delta_c$  are the strain response parameters for the conduction and valence bands. The activation of quantum emitters can be explained following Fig.~\ref{strain}. The dip in the deformation potential (Fig.~\ref{strain}(a)) increases the electron density around the truncated nanocone edge (Fig.~\ref{strain} (c)), thus assisting in quantum emitter activation. Fig.~\ref{strain}(c) shows the ratio of charge densities with ($d$) and without ($d_0$) strain. The charge density is calculated using the Boltzmann distribution from the deformation potential via the formula: $d=d_0\exp{(-V_{cb}/k_BT)}$ \cite{Proscia2018}. 

FDTD simulations performed in Ansys Lumerical show plasmonic enhancement of $\sim$25 at the top edges of the gold truncated nanocone for an excitation wavelength of 532 nm. However, since there is no spacer layer atop our truncated nanocones, there is a possibility of charge transfer and emission quenching due to proximity to the plasmonic cone. Thus, there could be a complicated interplay between strain and plasmonic effects which led to the observation of significant activated emission over most of the hBN/Au regions in Fig.~\ref{pl-color}. A careful study isolating the two phenomena will be performed in a subsequent work.

\section{Conclusion}
In this work, we fabricated gold truncated nanocone structures using e-beam lithography technique and successfully integrated hBN on them. We measured the emission from both samples: hBN on a planar Si/SiO$_2$ substrate and hBN integrated with the specifically engineered truncated nanocone structures. Our results indicate that hBN exhibits emission after integration with the engineered truncated nanocone structures. The activation of the emission in hBN emphasizes the significant impact of the truncated nanocone structures, validating the light-matter interaction in the hybrid system. Further, we demonstrated the quantum behavior of the emission. We revealed a negative impact from gold due to its fluorescence on the quantum emission characteristics. This underscores the need for a careful inspection of external influences during the exploration of the quantum behavior of such integrated systems. Finally, we presented strain calculations and electromagnetic field enhancement simulations as possible mechanisms for the emission activation. This study will be useful for the development of deterministic and tunable single photonic sources in two dimensional materials and their integration with plasmonic platforms.

\emph{Acknowledgement--} A.K.S acknowledges financial support from Industrial Research and Consultancy Center (IRCC) IIT Bombay. B.K. acknowledges support from Prime Minister’s research fellowship (PMRF), Government of India. A.K. acknowledges funding support from the Department of Science and Technology via the grants:  DST/CRG/2022/001170, SB/S2/RJN-110/2017,ECR/2018/001485 and DST/NM/NS-2018/49. We also acknowledge the Industrial Research and Consultancy Center (IRCC); the Centre of Excellence in Nanoelectronics (CEN), IIT Bombay; Fundamental Optics, THz and Optical Nanostructures (FOTON) laboratory at TIFR-Colaba, Mumbai; and Sophisticated Analytical Instrument Facility (SAIF); Centre for Research in Nanotechnology and Science (CRNTS)  IIT Bombay .
\bibliography{ref.bib}
\end{document}